\documentclass[conference]{IEEEtran}

\usepackage{graphicx}        % Add support for graphical figures
\usepackage[numbers]{natbib} % Add more flexibility for citing references
\usepackage{booktabs}

\begin{document}
	
\title{Antifragile Electronic Warfare}
\author{\IEEEauthorblockN{Marc Lichtman}
\IEEEauthorblockA{Virginia Tech\\Blacksburg, Virginia, USA\\Email: marcll@vt.edu}}
\maketitle

\begin{abstract}
	This letter introduces the concept of antifragile electronic warfare (EW), which we define as the ability to allow a communications link to improve performance due to the presence of a jammer.  This concept should not be confused with jamming countermeasures (a.k.a. anti-jamming or electronic protection).  Rather, antifragile EW can be thought of as the next step beyond simply avoiding or mitigating jamming.  
	%After introducing the concept we narrow down the subset of jammers this concept can be applied to, and provide a brief example of an antifragile EW strategy.  
	%This letter acts as a prepublication for a more extensive peer-reviewed journal paper (the full paper does not appear due to IEEE prepublication policy).
\end{abstract}


Antifragility is a concept popularized by professor Nassim Nicholas Taleb, and a term he coined in his 2012 book, \textit{Antifragile} \cite{taleb2012antifragile}.  Antifragility refers to systems that increase in capability, resilience, or robustness as a result of mistakes, faults, attacks, or failures \cite{taleb2012antifragile}.  As Taleb explains in his book, antifragility is fundamentally different from the concepts of resiliency (i.e. the ability to recover from failure) and robustness (i.e. ability to resist failure).  

Electronic warfare (EW) refers to any action involving the use of electromagnetic energy to control the electromagnetic spectrum.   EW is typically broken down into three subdivisions: electronic attack (EA), electronic protection (EP), and electronic support (ES).  EA and EP refer to jamming and anti-jamming respectively, while ES refers to searching, identifying, and locating jammers.  While EW encompasses wireless communications, radar, and directed energy weapons, we limit our analysis to wireless communications.

We propose adding a fourth subdivision to EW, called Antifragile Electronic Warfare (AEW), in which the communication system being jammed increases in performance as a result of the jamming attack.  AEW can be considered as an extension to EP or, rather, going a level beyond EP.  In a typical EP scheme, the radio performs waveform adaption (e.g. using cognitive radio techniques) in order to remain operable under jamming, with or without sacrificing performance.  However, in AEW, the radio is able to exploit a jamming attack to achieve higher effectiveness/performance as a result of jamming (an example is given in Table \ref{example_table}).

We will now investigate how to achieve a communications gain due to the presence of a jammer, and the conditions under which it is feasible.  Basic jammers, such as barrage or random jammers, operate as programmed and do not change their behavior.  If the communications system has no influence on the jammer, then it is impossible to cause the jammer to transmit desired information, or perform any other action that could lead to a communications gain.  There is a subcategory of jammers that react to sensed energy or signals, known as reactive jammers \cite{xu2005feasibility}.  These jammers must have receiving capability, and the most common form of reactive jamming is repeater jamming (a.k.a. digital RF memory or DRFM jamming) in which the jammer repeats the signal it receives.  In terms of information theory, if the transmitted signal $X$ is jammed by signal $J$, a communications gain is only possible if the mutual information between the two is greater than zero, denoted as $I(X;J)>0$. 

One way to improve capability of communications is to increase the amount of bits transferred between radios for a given amount of energy.  If a radio can transmit in a manner that causes a reactive jammer to relay its information without degrading the data or the original signal, the radio can harness the jammer's higher transmit power. In essence, one could create a customized waveform and information alphabet by exploiting the characteristics of the jammer's response. By providing this second (redundant) data stream to the receiver, the error rate can be reduced (increasing the amount of bits transferred to the receiver), and thus provide an AEW advantage.  For example, if the jammer transmits noise on all channels that it believes contain energy, then the communications system could begin using wideband multi-frequency FSK across selected multiple channels (to create an communications alphabet), in such a way that the intended communication signal and the jamming signal are receiving orthogonally and soft-combined after being separately demodulated.  The specific hopping rate would be a function of the reactive jammer's delay and channels covered, and would have to take into account any timing jitter.  Hence, the reactive jammer's transmitted waveform can be treated as a redundant version of the signal, which can potentially improve the bit error rate without requiring the communications system to transmit additional energy. The jammer's effect is similar to introducing multipath propagation, with relatively long delay spreads and there are numerous signal processing techniques that can be used to exploit multipath (delayed transmission). For example, the AEW receiver could implement some sort of maximum ratio combining after proper channel estimation and equalization for each signal path (receive diversity). This is just a simple example; more approaches to AEW are envisioned and will be analyzed in future research. 

\begin{figure}
	\centering
	\includegraphics[width=6in]{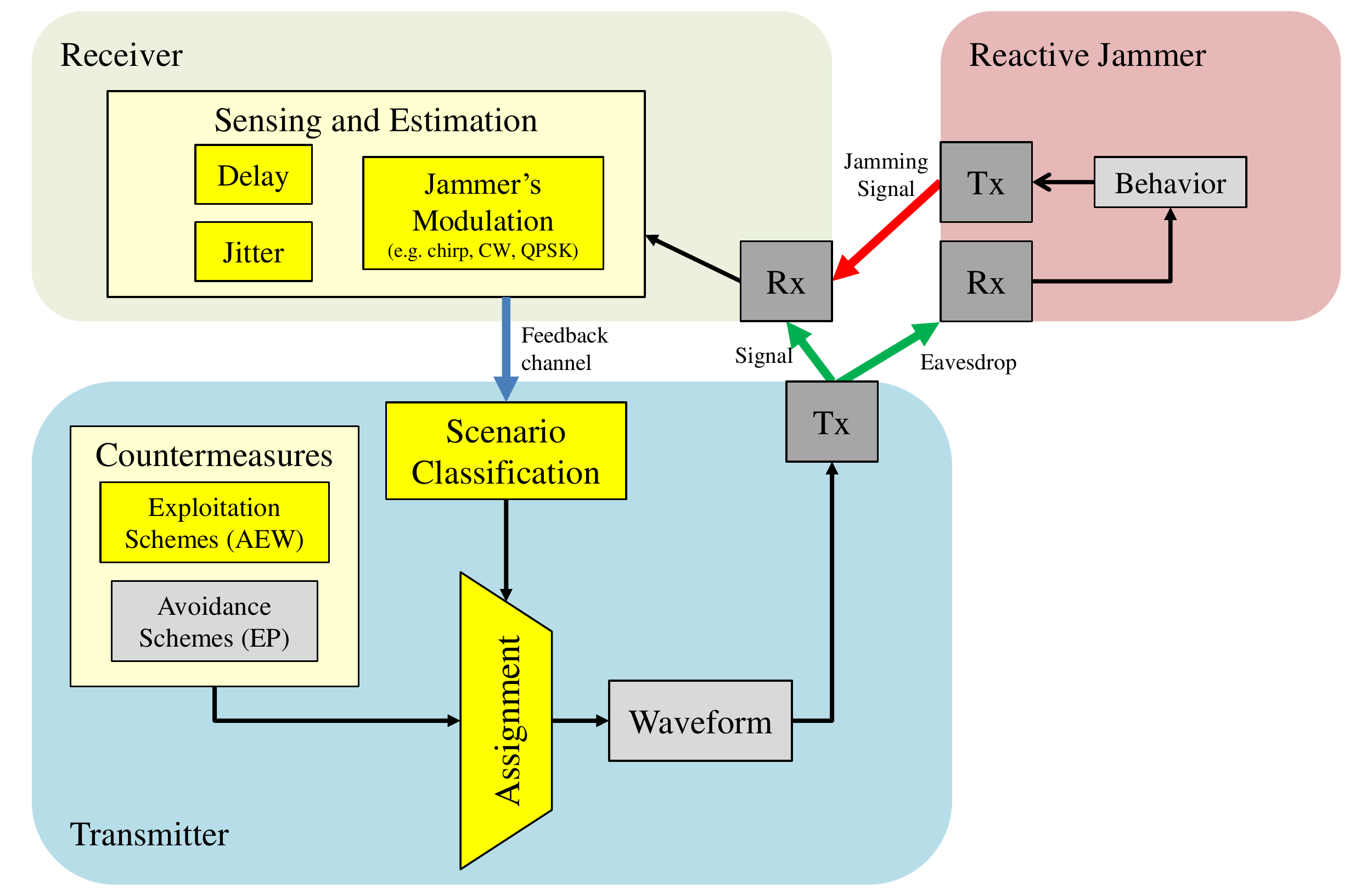}
	\caption{The components of the proposed AEW system (highlighted in yellow) and how they fit into a communications system}
	\label{main}
\end{figure}

Simply knowing that a reactive jammer is present is not enough to perform AEW.  A functioning AEW strategy is based on the scenario at hand; i.e. the type of communications link, the reactive jammer's characteristics, and the delays/jitters involved.  Therefore, an effective strategy involves implementation of a series of different schemes that exploit the jammer and a classifier that can identify the scenario at hand and estimate parameters associated with it.  Lastly, it must incorporate an engine that can assign the most effective AEW scheme to the given scenario.  Figure \ref{main} illustrates the components of the proposed AEW system (highlighted in yellow) and how they fit into a communications system. 

As the sophistication of communications systems and jammers increases, reactive jamming will likely become a bigger threat in military and other mission-critical domains.  Therefore, it may be worthwhile to consider incorporating an antifragile component into state of the art protected radios.  As part of future research we will continue to develop the concept of AEW.

\begin{table}[h]
	\centering
	\caption{A simple example of the difference between mitigation and antifragility}
	\begin{tabular}{cc}
		\toprule
		\textbf{Scenario} & \textbf{Example Throughput} \\
		\midrule 
		Non-jammed (baseline) & 1 Mbps \\ 
		Jammed without built-in countermeasures & 0.2 Mbps \\ 
		Jammed with built-in countermeasures & 0.9 Mbps \\ 
		Jammed with built-in Antifragility (AEW) & 1.1 Mbps \\ 
		\bottomrule
	\end{tabular} 
	\label{example_table}
\end{table}

\bibliographystyle{IEEEtranN}
\bibliography{references}

\onecolumn

\end{document}